\documentclass[12pt]{article}
\usepackage{graphicx}

\usepackage{array}
\usepackage{psfrag}
\usepackage{braket}
\usepackage{subfig}
\usepackage[usenames,dvipsnames]{color}
\usepackage[hyperfootnotes=false,colorlinks=true, linkcolor=BrickRed, citecolor=Blue, urlcolor=Blue, filecolor=Blue]{hyperref}
\usepackage{epsfig,amsmath,amssymb,fancyhdr}   
\usepackage[normalem]{ulem}
\usepackage{amsmath}
\usepackage{slashed}
\usepackage{mathtools}
\usepackage[titletoc]{appendix}
\usepackage{datetime}
\usepackage{alltt} 
\usepackage{esvect}
\usepackage{graphicx}
\usepackage{bookmark}

\makeatletter
\def\@xfootnote[#1]{%
  \protected@xdef\@thefnmark{#1}%
  \@footnotemark\@footnotetext}
\makeatother

\newcommand{\remove}[1]{}

\DeclarePairedDelimiter{\evdel}{\langle}{\rangle}
\newcommand{\ev}{\evdel} % For expectation values

\newcommand\pubnumber{NuPhys2016-Olivares-Del Campo}
\newcommand\pubdate{\today}

\def\support{\footnote[*]{$\,$ Poster presented at NuPhys2016. This work has been supported
by  the  European  Research  Council  under  ERC  Grant
NuMass  (FP7-IDEAS-ERC  ERC-CG  617143).}}
\def\Title#1{\begin{center} {\Large #1 } \end{center}}
\def\Author#1{\begin{center}{ \sc #1} \end{center}}
\def\Address#1{\begin{center}{ \it #1} \end{center}}

\newcommand\pubblock{\rightline{\begin{tabular}{l} \pubnumber\\
         \pubdate  \end{tabular}}}
\newenvironment{Abstract}{\begin{quotation}  }{\end{quotation}}
\newenvironment{Presented}{\begin{quotation} \begin{center} 
             PRESENTED AT\end{center}\bigskip 
      \begin{center}\begin{large}}{\end{large}\end{center} \end{quotation}}

\def\beq{\begin{equation}}
\def\eeq#1{\label{#1}\end{equation}}
\def\eeqn{\end{equation}}

\def\beqa{\begin{eqnarray}}
\def\eeqa#1{\label{#1}\end{eqnarray}}
\def\eeqan{\end{eqnarray}}

\let\bar=\overbar

\def\Dslash{\not{\hbox{\kern-4pt $D$}}}
\def\dslash{\not{\hbox{\kern-2pt $\del$}}}

\def\msb{{\bar{\ssstyle M \kern -1pt S}}}

\begin{document}
\begin{titlepage}
\pubblock

\vfill
\Title{Phenomenology of a Neutrino-DM Coupling: The Scalar Case}
\vspace{1cm}
\Author{C\'eline B\oe hm$^{1,2}$, \textbf{Andres Olivares-Del Campo}$^{1}$ \support, Sergio Palomares-Ruiz$^{3}$, Silvia Pascoli$^{1}$ }
\Address{$^1\;$Institute for Particle Physics Phenomenology, Durham University, South Road, Durham, DH1 3LE, United Kingdom}
\Address{$^2\;$LAPTH, U. de Savoie, CNRS,  BP 110, 74941 Annecy-Le-Vieux, France}
\Address{$^3\;$Instituto  de  Fisica  Corpuscular  (IFIC),  CSIC-Universitat  de  Valencia,
Apartado  de  Correos  22085,  E-46071,  Spain}
\vfill
\begin{Abstract}
Dark matter (DM) and neutrinos are the two most compelling pieces of evidence of new physics beyond the Standard Model of Particle Physics but these are often treated as belonging to two different sectors. Yet DM-neutrino interactions are known to have cosmological consequences. Here, we study the scenario of a scalar DM candidate coupled to left-handed neutrinos via a Dirac mediator. We determine the mass of a DM candidate that yields the right DM relic abundance in a thermal scenario and it is consistent with large-scale structure formation. In order to satisfy both constraints, a complex DM candidate should have a mass larger than $8.14$ keV while the mass of a real DM candidate should be above $18.1$ eV, independently of the value of the DM-neutrino coupling.
\end{Abstract}
\vfill
\begin{Presented}

NuPhys2016, Prospects in Neutrino Physics

Barbican Centre, London, UK,  December 12--14, 2016

\end{Presented}
\vfill
\end{titlepage}
\def\thefootnote{\fnsymbol{footnote}}
\setcounter{footnote}{0}

\newpage

%%%%%%%%%%%%%%%%%%%%%
\section{Introduction}
%%%%%%%%%%%%%%%%%%%%
The discovery of neutrino masses together with the presence of dark matter (DM) strongly suggest the existence of new physics beyond the Standard Model of Particle Physics. Consequently, models were proposed to explain  both phenomena (and in particular the relic density and neutrino masses) in a minimalistic way \cite{CelineMa}. Such attempts generally consider a DM-neutrino interaction term which can lead to a rich phenomenology in the Early Universe. Furthermore, such interactions introduce the possibility of detecting neutrinos that are produced from DM self-annihilation at neutrino detectors on Earth \cite{Sergio1}. So far only a limited number of models have been considered in the literature for dark matter-neutrino interactions \cite{CelineFayet}. However, given that DM particles have not been found yet and that the mechanism by which neutrinos acquire a mass remains unsettled, it is worth investigating a larger number of scenarios and examine whether they are compatible with known constraints.

%%While some results are straight forward, there are some subtleties depending of the nature of the scalar candidate (complex or real) or the nature of the fermionic mediator (Dirac or Majorana) that will show significant differences and allow us to distinguish between scenarios. 

In this paper, we will consider the scenario of a scalar DM candidate coupled to left-handed neutrinos via a Dirac fermion\footnote{$\,$This is done as a proof of concept and we leave the full discussion of all possible scenarios consistent with Lorentz invariance where a DM candidate can interact with neutrinos to a future paper that will be released shortly.} so that:
\begin{equation}
\mathcal{L}_{\rm{int}} \supset -\,g\,\chi \, \bar{N}_R \, \nu_L   \ +\ \mathrm{h.c.}~, \label{lag}
\end{equation}
where $N$ is the Dirac mediator and $\chi$ the DM candidate. Such coupling can arise by introducing a Dirac $SU(2)$ doublet like in supersymmetric models \cite{CelineFayet}. If on the other hand, $N$ is a singlet, the coupling can be generated in Inert Doublet models where the scalar $\chi$ belongs to an $SU(2)$ doublet. Since the aim of this paper is to study the cosmological implications of the coupling $g$, we take the DM and mediator masses as free parameters and we don't discuss any model specific bounds. %Alternatively, the scalar can be a singlet with the Dirac fermion belonging to a doublet under $SU(2)$ like in $D>6$ theories \cite{6D}.

The paper is structured as follows: In the next section we will briefly review the different experimental signatures considered to test our scenario and we will discuss the results in Section 3. Finally, we will conclude in Section 4.

%{Mangano:2006mp,Wilkinson:2014ksa} and the structure of galaxies \cite{Boehm:2000gq, Boehm:2004th,Boehm:2014vja, Schewtschenko:2014fca, Schewtschenko:2015rno}.  

\section{Cosmological signatures}
%%%%%%%%%%%%%%%%%%%%
A DM-neutrino interaction induces processes such as the annihilation of DM to neutrinos and the elastic scattering between neutrinos and DM particles. If this is the dominant annihilation channel for thermal freeze-out, the thermally averaged annihilation cross section of DM to neutrinos will set the amount of DM that we observe today (i.e. the relic density), for which $\ev{\sigma v_{\rm{r}}}_{\text{Th}}\simeq 3 \times 10^{-26} \;\rm{cm}^3\ \rm{s}^{-1}$ is required. As we impose the DM candidate's relic density to be smaller or equal to the observed abundance, $\Omega_{\rm{DM}}h^2=0.1188$ \cite{Eleonora}, we obtain a lower bound on the strength of the DM-neutrino interaction.
% where $v_{CM}\sim \frac{1}{3}\,\rm{c}$ in the early universe.

It has also been shown that the elastic scattering between neutrinos and DM can lead to a suppression of small-scale structures in the Universe (known as collisional damping) \cite{Wilkinson:2014ksa}, since it allows for DM to be in equilibrium with neutrinos even after chemical decoupling. Therefore, the DM takes longer to free-stream and could lead to a further suppression of such structures. %\cite{Boehm:2000gq,Boehm:2001hm,Boehm:2004th,Boehm:2014vja,Schewtschenko:2014fca,Schewtschenko:2015rno}.
By confronting the large-scale structure predictions to observations, the relevant constraint for the scenario considered is \cite{Wilkinson:2014ksa} 
\begin{equation}
 \sigma_{\rm{el}}< 10^{-48} \ \left(\frac{m_{\rm{DM}}}{\rm{MeV}}\right) \ \left(\frac{T_0}{2.35 \times 10^{-4} \ \rm{eV}}\right)^2 \  \rm{cm^2}, \label{col}  
\end{equation}
for an energy dependent cross section, with $T_0$ the neutrino temperature today.

Finally, in the presence of DM-neutrino interactions, one can search for a flux of neutrinos and anti-neutrinos produced from DM annihilation at rest in regions with a high DM density like the Milky Way. Such a flux would be monochromatic since each neutrino will carry an energy equal to the DM mass and could be detected at neutrino detectors on Earth. 
For this scenario, the relevant constraints in the total annihilation cross section of DM to neutrinos can be set using the Super-Kamiokande (SK) Phase I data from the supernova relic neutrino search following the analysis done in \cite{Sergio1}. Nevertheless, the results including SK Phases I-III are expected to be similar \cite{Bays1}. 

%It is worth mentioning that, the coupling in Eq. \ref{lag} can also induce a neutrino mass term in a specific model like 

%where $\sigma_{\text{det}}$ is the detector cross section and is evaluated at $E_\nu=m_{\rm{DM}}$. $\phi$ is the total flux of neutrinos or anti-neutrinos, $N_{\text{target}}$ is the total number of particles in the detector, t is the exposure time and $\epsilon$ is the energy efficiency.

\renewcommand{\arraystretch}{2}
\begin{table}[!b]
\centering
\begin{tabular}{  c |  c  c c  }
 & \hspace{0.2cm} \textbf{Complex DM} & \hspace{0.2cm} \textbf{Real DM} \\
\hline 
$\ev{\sigma v_{\rm{r}}} \propto$ & $g^4v^2_{\rm{CM}}\frac{m^2_{\rm{DM}}}{(m^2_{\rm{DM}}+m^2_{\rm{N}})^2}$ & $g^4v^4_{\rm{CM}}\frac{m^6_{\rm{DM}}}{(m^2_{\rm{DM}}+m^2_{\rm{N}})^4}$  \\ 
$~~\sigma_{\rm{el}} \,\, \propto$ & $g^4 E^2_\nu \frac{1}{(m^2_{\rm{DM}}-m^2_{\rm{N}})^2}$ & $  g^4 E^4_\nu \frac{m^2_{\rm{DM}}}{(m^2_{\rm{DM}}-m^2_{\rm{N}})^4}$  \\  
\hline 
\end{tabular}
\caption{Relevant terms of the expressions for the annihilation and the elastic scattering cross sections for a complex and a real DM candidate when $m_{\rm{DM}}\neq m_{\rm{N}}$.}
\label{Table1}
\end{table}

\section{Results}
\begin{figure}[!t]
\minipage{0.5\textwidth}
  \includegraphics[width=8cm,height=6cm,keepaspectratio]{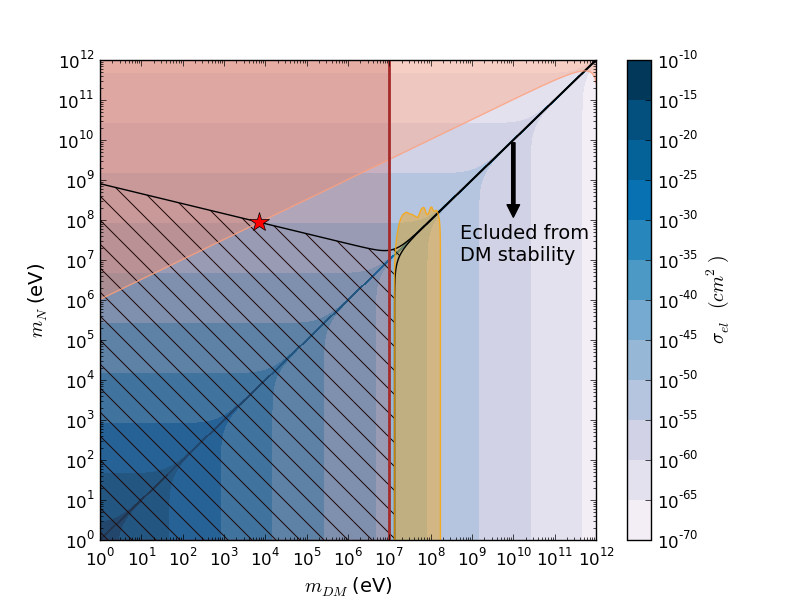}

\endminipage\hspace{-0cm}
\minipage{0.5\textwidth}
  \includegraphics[width=8cm,height=6cm,keepaspectratio]{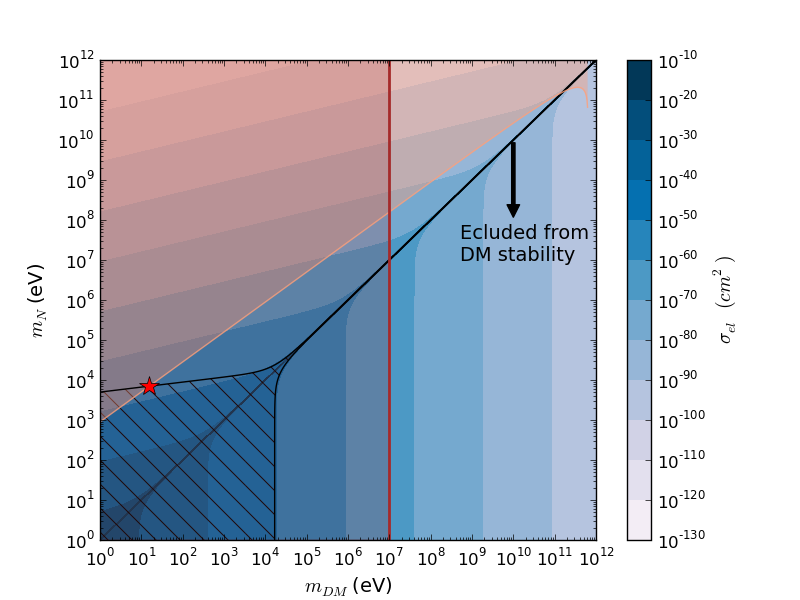}

\endminipage
\vspace{-0.0cm}
\caption{Relevant parameter space in the $m_{\rm{N}}-m_{\rm{DM}}$ plane with $g=1$ for complex DM (left) and real DM (right). The light red and orange regions correspond to DM overproduction and the 90$\%$ C.L. bound from the SK search using $v_{\rm{CM}}=\frac{1}{3}$ c and $v_{\rm{CM}}=10^{-3}$ c respectively. The dashed area represents the excluded region from collisional damping while the brown vertical line refers to the lower bound from Planck effective number of neutrino species measurement. The red star is the point from which both the relic density and the collisional damping constraints are satisfied.
}
\label{ComplexDMDiracMed}
\end{figure}
\vspace{-0.2cm}

We compute the thermal annihilation cross section to left-handed neutrinos and the elastic scattering cross section between DM and left-handed neutrinos in the limit $m_\nu \rightarrow 0$, which is summarized in Table \ref{Table1}. Fig. \ref{ComplexDMDiracMed} shows the allowed parameter space considering the relevant constraints. The colored contours correspond to different values for the elastic scattering cross section. It is worth noting that, for the scalar to be a viable DM candidate, $m_{\rm{N}}>m_{\rm{DM}}$ so that the DM remains stable. Consequently, the parameter space below the diagonal in Fig. \ref{ComplexDMDiracMed} is excluded.

%\begin{equation}
%\ev{\sigma \text{v}_{\rm{r}}}, \ \ (\sigma_{\rm{el}})\propto\begin{cases}  g^4v^2_{\rm{CM}}\frac{m^2_{\rm{DM}}}{(m^2_{\rm{DM}}+m^2_{\rm{N}})^2},\ \ \Big{(}  g^4 E^2_\nu \frac{1}{(m^2_{\rm{DM}}-m^2_{\rm{N}})^2}\Big{)} & \text{Complex DM}\\ g^4v^4_{\rm{CM}}\frac{m^6_{\rm{DM}}}{(m^2_{\rm{DM}}+m^2_{\rm{N}})^4}, \ \ \Big{(}  g^4 E^4_\nu \frac{m^2_{\rm{DM}}}{(m^2_{\rm{DM}}-m^2_{\rm{N}})^4}\Big{)} & \text{Real DM}.
%  \end{cases}\label{cross}
%\end{equation}

The cross sections for real DM are more suppressed due to the $v^4_{\rm{CM}}$ and $E^4_\nu$ dependence, which in turn translates into weaker bounds as can be seen in Fig. \ref{ComplexDMDiracMed}. However, in the degenerate regime (i.e. $m_{\rm{DM}}\sim m_{\rm{N}}$) $\sigma_{\rm{el}}\propto g^4/m^2_{\rm{DM}}$ which shows as an enhancement of the elastic cross section along the diagonal in Fig.~\ref{ComplexDMDiracMed}. Moreover, the p and d-wave dependence of the annihilation cross section implies that $\ev{\sigma v_{\rm{r}}}$ today is very small since $v_{\rm{CM}}\sim 10^{-3}$ c today. Consequently, neutrino detectors do not provide any bounds for large DM masses and the constraints from the SK analysis only apply to the complex DM scenario as it is less suppressed.
% But this region is still constrained from LSS. 

% The different structures occur due to a cancellation between the t and u (s) channels in the calculation of the annihilation (elastic scattering) cross section for a real DM candidate.

%Furthermore, for illustration purposes, we assume that the coupling $g\sim 1$. This requirement can be relaxed and it will affect both, the elastic and annihilation cross sections equal as both are $\propto g^4$. Therefore, the final results are independent of the coupling as it is discussed below.  

As can be seen from Table \ref{Table1}, the coupling $g$ enters with the same power in the annihilation and elastic scattering cross sections. Hence, we can impose $\ev{\sigma v_{\rm{r}}} \sim 3 \times 10^{-26}\;\rm{cm}^3 \ \rm{s}^{-1}$ to get a coupling independent expression for the elastic scattering, which can be compared to the collisional damping constraint in Eq. \ref{col}:
\begin{equation}
\sigma_{el} \simeq  5.41 \times 10^{-55} \ \left(\frac{T_0}{2.35\times 10^{-4} \ \rm{eV}} \right)^{2}  \ \left(\frac{m_{\rm{DM}}}{\rm{MeV}}\right)^{-2} \  \left( \frac{\ev{\sigma v_{\rm{r}}}}{3 \times 10^{-26} \ \rm{cm^3/s}}\right)\ \rm{cm^2}, \label{el1COMBINED}
\end{equation}
for complex DM when $m_{\rm{N}}>m_{\rm{DM}}$ and
\begin{equation}
\sigma_{el} \simeq  1.96 \times 10^{-72} \ \left(\frac{T_0}{2.35\times 10^{-4} \ \rm{eV}} \right)^{4} \  \left(\frac{m_{\rm{DM}}}{\rm{MeV}}\right)^{-4} \ \left( \frac{\ev{\sigma v_{\rm{r}}}}{3 \times 10^{-26} \ \rm{cm^3/s}}\right)\ \rm{cm^2}, \label{el1COMBINED}
\end{equation}
for real DM when $m_{\rm{N}}>m_{\rm{DM}}$, where we have assumed $E_\nu \sim T_0$ today. Therefore, if we want to satisfy the collisional damping and relic density constraints in the limit $m_{\rm{N}}>m_{\rm{DM}}$, we require DM masses larger than $8.14 $ keV (18.1 eV) and mediator masses larger than 87.7 MeV (6.97 keV) for complex (real) DM for any coupling $g$ (red star in Fig. \ref{ComplexDMDiracMed}). Nevertheless, it has been shown that the effective number of neutrino species when DM is in thermal equilibrium with neutrinos is only consistent with Planck measurements for $m_{\rm{DM}}\gtrsim 10$ MeV for both scenarios \cite{CelineNeff}. This corresponds to $m_{\rm{N}} \gtrsim 3.55$ GeV for complex DM and $m_{\rm{N}} \gtrsim 0.14$ GeV for real DM and imposes stronger bounds than the collisional damping constraint.

%in the complex DM scenario\footnote{$\,$Note that there is no constraint for real DM due to the d-wave dependence of $\ev{\sigma\rm{v}_{\rm{r}}}$.}.

This analysis shows that there are subtleties when analyzing the different scenarios that can lead to a very distinct phenomenology. This is also the case when, for example, one considers a Majorana mediator instead of a Dirac mediator since this could produce Lepton Number Violating processes such as $\chi \bar{\chi} \rightarrow \nu_{L}\nu_{L}$.

\section{Conclusion}
The study of neutrino-DM interactions is a powerful tool to constraint the masses of the DM and its mediator since it provides a variety of cosmological observables to contrast with the theoretical predictions. Furthermore, the complementarity of such observables with model-specific predictions allows us to understand better what the nature of the DM particle could be. Here we have only discussed the scenario of scalar DM coupled to neutrinos via a Dirac mediator, but a full study of all the possible scenarios will help us to determine the allowed values of the parameter space for DM candidates and mediators of different spins.

\bibliography{NuPhysProceedings}{}
\bibliographystyle{physRev}

%\begin{thebibliography}{99}

%\bibliography{Biblio.bib}
%%
%%  bibliographic items can be constructed using the LaTeX format in SPIRES:
%%    see    http://www.slac.stanford.edu/spires/hep/latex.html
%%  SPIRES will also supply the CITATION line information; please include it.
%%
%%CITATION = HEP-PH/0612228;%%"}
%\bibitem{Mesmer}
%F. A. Mesmer, Proc. Wien. Acad. Sci. {\bf 13}, 1564, 1593 (1762).
%%CITATION = PWASA,13,1564;%%

%\end{thebibliography}

\end{document}